\documentclass[aps,prl,twocolumn,preprintnumbers,
superscriptaddress,floatfix,nofootinbib]{revtex4}

\usepackage{multirow, graphicx,amssymb,url,mathrsfs,amsmath}
\usepackage{eucal,wrapfig,boxedminipage,setspace,subfigure}
\usepackage{amsxtra,amstext,latexsym,dsfont}

\def\IR{{\hbox{{\rm I}\kern-.2em\hbox{\rm R}}}}
\def\IB{{\hbox{{\rm I}\kern-.2em\hbox{\rm B}}}}
\def\IN{{\hbox{{\rm I}\kern-.2em\hbox{\rm N}}}}
\def\IC{\,\,{\hbox{{\rm I}\kern-.59em\hbox{\bf C}}}}
\def\IZ{{\hbox{{\rm Z}\kern-.4em\hbox{\rm Z}}}}
\def\IP{{\hbox{{\rm I}\kern-.2em\hbox{\rm P}}}}
\def\IH{{\hbox{{\rm I}\kern-.4em\hbox{\rm H}}}}
\def\ID{{\hbox{{\rm I}\kern-.2em\hbox{\rm D}}}}

\def\be{\begin{equation}}
\def\ee{\end{equation}}
\def\ba{\begin{eqnarray}}
\def\ea{\end{eqnarray}}

\def\ea{{\it et al}. }

\newcommand{\beq}{\begin{equation}}
\newcommand{\eeq}{\end{equation}}
\newcommand{\bea}{\begin{eqnarray}}
\newcommand{\eea}{\end{eqnarray}}

\begin{document}

\voffset 1cm

\newcommand\sect[1]{\emph{#1}---}

\pagestyle{empty}

\preprint{
\begin{minipage}[t]{3in}
\begin{flushright} SHEP-13-04
\\[30pt]
\hphantom{.}
\end{flushright}
\end{minipage}
}

\title{Holographic Modelling of  a Light Techni-Dilaton}

\author{Nick Evans}
\email{evans@soton.ac.uk}
\affiliation{ STAG Research Centre, Department of Physics and Astronomy, University of
Southampton, Southampton, SO17 1BJ, UK}
\author{Kimmo Tuominen}
\email{kimmo.i.tuominen@jyu.fi}
\affiliation{Department of Physics, University of Jyv\"askyl\"a, P.O.Box 35, FIN-40014 Jyv\"askyl\"a, Finland \\
and Helsinki Institute of Physics, P.O.Box 64, FIN-00014 University of Helsinki, Finland}

\begin{abstract}
We present a 
simplified holographic model of chiral symmetry breaking in gauge theory. The chiral condensate is represented by a single scalar field in AdS, with the gauge dynamics input through radial dependence of its mass, representing the running of the anomalous dimension
of the $\bar{q} q$ operator. We discuss simple examples 
%display the key elements of 
of the chiral transition out of the conformal window 
%- a BKT/Miransky scaling transition 
when the infrared value of the anomalous dimension, $\gamma_m$, is tuned to one (equivalently the AdS-scalar mass squared is tuned to the Breitenlohner-Freedman bound of -4). The output of the model are the masses of the $\bar{q} q$ scalar meson bound states. We show in an explicit example that if the gradient of the running of the anomalous dimension falls to zero at the scale where the BF bound violation occurs, so that  the
theory becomes near conformal, then the theory possesses a techni-dilaton state that is parametrically lighter than the dynamically generated quark mass. Indeed the full spectrum of excited meson states also become light (relative to the techni-quark mass) as they approach a conformal spectrum. 

\noindent

\end{abstract}

\maketitle

\newpage

There has been considerable theoretical interest in gauge theories that possess conformal windows (non-trivial infrared (IR) conformal fixed points in some range of quark flavours $N_f$) \cite{Caswell:1974gg}-\cite{Miransky:1996pd}. For a theory with quarks in the fundamental representation asymptotic freedom sets in when $N_f < 11/2 N_c$. Immediately below that point, at least at large $N_c$, the two loop beta function enforces a perturbative infra-red (IR) fixed point 
\cite{Caswell:1974gg,Banks:1981nn}.  The fixed point behaviour is expected to persist into the non-perturbative regime as $N_f$ is further reduced \cite{Appelquist:1996dq}. At some critical value of the number of flavours, $N_f^c$, the coupling is expected to be strong enough to trigger chiral symmetry breaking by the formation of a quark anti-quark condensate. 
The critical value, $N_f^c$, i.e. the lower boundary of the conformal window, can be estimated in a variety of ways \cite{Appelquist:1996dq}-\cite{Jarvinen:2011qe}. 
Different semi-analytic methods typically yield the chiral transition to occur below $N_f^c\simeq  4N_c$ for fundamental fermion flavours. 

The chiral phase transition at the lower edge of the window, may give way to a regime of walking dynamics directly below $N_f^c$
\cite{Holdom:1981rm}.  For walking theories, there is expected to be a long energy range in which the coupling barely runs before tripping through the critical coupling value in the deep IR. Such theories display
a tuned gap between the value of the quark condensate and the pion decay constant $f_\pi$. These theories are 
of interest phenomenologically for technicolor models of electroweak symmetry breaking \cite{Weinberg:1975gm, Susskind:1978ms} since  the quark bilinear has a significant anomalous dimension, $\gamma_m$, over a large energy regime. This in turn leads to the suppression of flavour changing neutral currents in extended technicolor models \cite{Eichten:1979ah}, and decouples flavour physics from the electroweak scale \cite{Holdom:1981rm}.
The walking dynamics may also suppress the contributions of the techni-quarks to the electroweak oblique corrections, in particular to the S parameter \cite{Sundrum:1991rf,Appelquist:1998xf}.  
It is believed that in walking theories the intrinsic scale falls towards zero exponentially with $N_f^c-N_f$ as the conformal window is approached from below (Miransky scaling \cite{Miransky:1996pd}). 
 
There has also been speculation that walking theories may possess a parametrically light bound state, a pseudo-Goldstone boson of the breaking of dilatation symmetry by the quark condensate \cite{Yamawaki:1985zg}-\cite{Appelquist:2010gy}.
This
is of particular interest given the recent discovery of a light ``higgs-like'' state at the LHC \cite{atlas:2012gk,cms:2012gu}. See \cite{Matsuzaki:2012xx}-\cite{Matsuzaki:2012vc} for recent discussions.

\begin{figure}[]
\centering
%\hspace{-2mm}
\includegraphics[width=6.5cm]{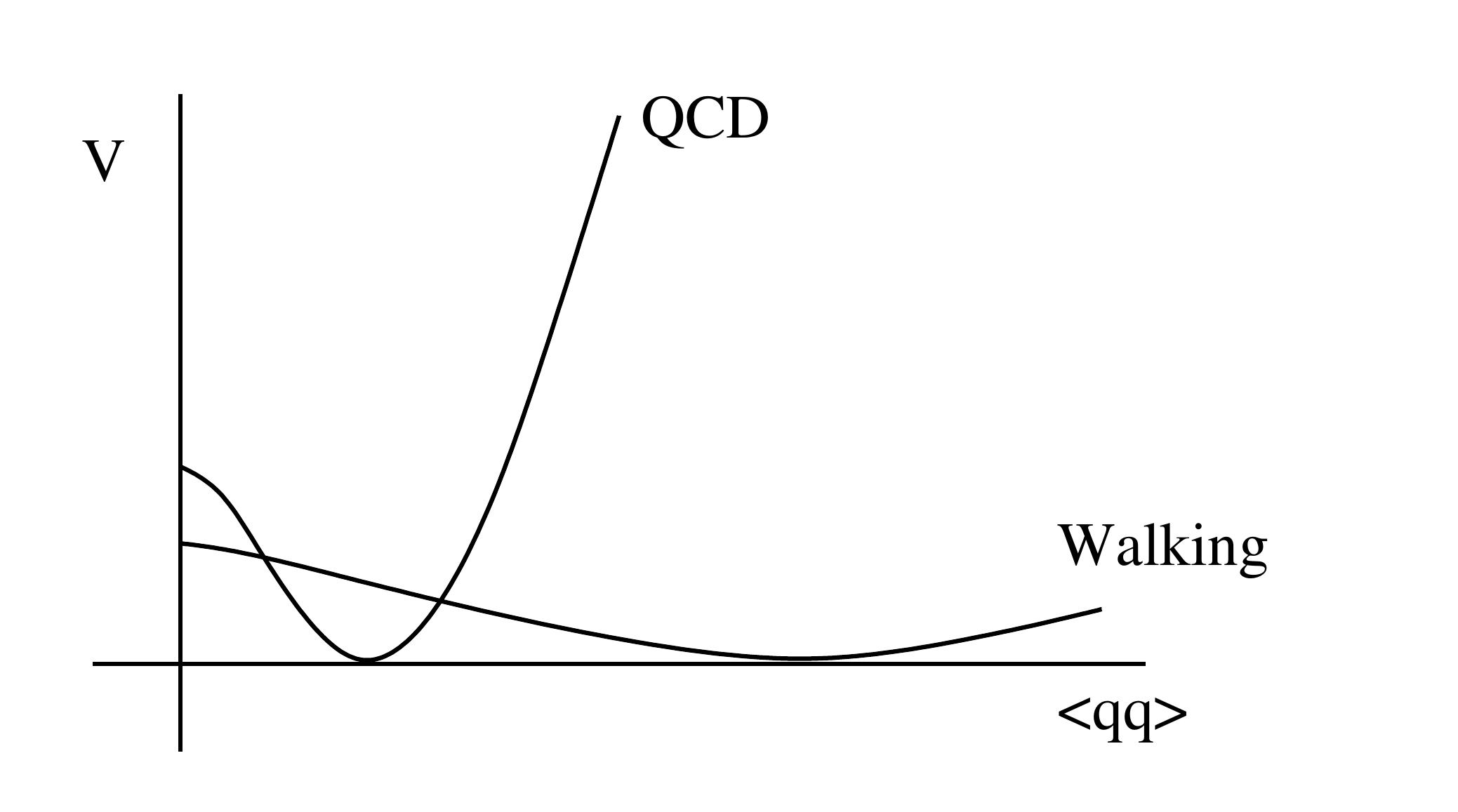}
\caption{Sketch of the effective potential for the quark condensate in QCD and a walking theory. The flatness of the potential for the walking case implies a light $\sigma$ particle.}
\label{potshape}
\end{figure}

The argument for the existence of a light techni-dilaton state seems simple: At the critical value the running coupling has an extremely long running regime infinitesimally close to the critical coupling. When it finally crosses the critical coupling value the condensate forms with a value determined by the renormalization scale at which the critical coupling was met. However, the theory is so close to conformal that preference on a particular scale is very modest, and a slightly larger or smaller condensate value would be almost as energetically likely. The potential has become extremely flat (see Fig. \ref{potshape}) in the $\sigma$ mode direction associated with the magnitude of the condensate, and a Goldstone like $\sigma$ scalar should be present in the spectrum.\footnote{Note this flattening of the potential in a walking theory is very much related to the arguments in \cite{Evans:2010tf,Evans:2012jx} that the potential is suitable for generating inflation which also requires a flat potential between zero condensate and the true vacuum.}. 

In this paper we will explore a simple holographic \cite{Maldacena:1997re,Witten:1998qj,Gubser:1998bc} model for the chiral symmetry breaking transition at the edge of the conformal window. Such holographic models have been recently addressed in the literature
\cite{Jarvinen:2009fe}-\cite{Elander:2009pk}.
The crucial element of these models is that the quark bilinear is dual to a scalar in the AdS-like bulk, and chiral symmetry breaking is triggered when its mass in the infrared falls through the Breitenlohner-Freedman (BF) bound of $m^2=-4$ \cite{Breitenlohner:1982jf}.  Using the naive AdS/CFT dictionary, appropriate in such a near conformal setting, $m^2= \Delta(\Delta-4)$ \cite{Witten:1998qj}, where $\Delta$ is the dimension of the operator, 
this corresponds to $\gamma_m=1$ in the gauge theory. 
%%KT: Revised the text below on BKT:
In a class of models, the resulting transition is of the holographic BKT (Berezinskii - Kosterlitz - Thouless)  type \cite{Kaplan:2009kr,Jensen:2010ga,Evans:2010hi} showing the anticipated exponential (Miransky) scaling in $\Delta m^2$ as one moves from the critical point. Of course other possibilities exist; we will exhibit one model where the transition is of first order and another where BKT type transition occurs.
The models to date have attempted to predict the dynamics of the gauge
theory and the associated critical value of $N_f$. For example, the model in \cite{Alvares:2012kr} inputs a running gauge coupling and
then predicts the resultant running of the AdS scalar mass. The model in \cite{Jarvinen:2011qe} is even more sophisticated 
including back reaction of flavor degrees of freedom on the background geometry describing the gluodynamics. 
A downside of these models is that one can not easily dial the running of the  AdS scalar mass/$\gamma_m$ to study the effect on the spectrum. For this reason, here we will take a very simplistic and phenomenological approach. 
Although our model links naturally to the linearized Dirac-Born-Infeld (DBI) action of the D3/D7 system models \cite{Karch:2002sh,Babington:2003vm,Alvares:2012kr},
it is  simply a single scalar field in AdS space that describes the $\bar{q} q$ operator and its associated spectrum. 
We input the gauge dynamics by hand by simply  choosing the radial 
dependence of the quark mass which we interpret as the renormalizantion group (RG) flow of the dimension of the quark bilinear.
The model then lets us compute the scalar bound state spectrum. 

As already emphasized, in the dynamics of quasi conformal theories the main roles are played by explicit and spontaneous scale breaking. The explicit breaking of scale invariance is controlled by the nontrivial renormalization group running, while the spontaneous breaking is due to dynamical formation of the chiral condensate. To demonstrate the features of these effects, we consider two choices for $\Delta m^2$, both interpolating from zero in the ultraviolet (UV) to negative values in the infrared (IR). Our first choice has a single dimensionful parameter $\alpha$ which controls both the scale of symmetry breaking and the gradient of $\gamma_m$. Due to the simple structure, we are able to determine the dependence of the spectrum on $\alpha$ analytically.

Our second example is a two scale ansatz for $\Delta m^2$: one parameter fixes the scale where 
the BF bound is violated, and the other fixes the gradient of $\gamma_m$ at that scale. This model allows us to approach conformality for symmetry breaking at a fixed scale. Here we find that the lightest 
scalar quark bound state's mass versus the IR value of the techni-quark mass, i.e. $f_\pi$, indeed falls to zero as we approach conformality.   In addition, we find that the  higher radially excited states of the 
$\sigma$ are also falling to zero mass. We interpret this as necessary for the spectrum to 
complete a continuous conformal spectrum in
the conformal limit. Interestingly, as we approach the conformal limit the lightest scalar becomes light relative to the excited states suggesting an interpretation as a true techni-dilaton may be appropriate. 

We believe these analyses exhibit the general features relevant for the spectrum of a walking theory and makes the tuning involved to obtain a techni-dilaton in the 
spectrum very clear. 

\section{The Model}

Our model is inspired by the analysis of chiral symmetry breaking in the D3/probe-D7 gravity duals of \cite{Karch:2002sh,Babington:2003vm} - see \cite{Alvares:2012kr} for its particular application to the conformal window.  The position of the D7 brane in AdS$_5\times$S$^5$ space of radius $R$ is described by the  DBI action of the form
\begin{equation}
 S \sim \int d \rho e^\phi \rho^3 \sqrt{1 + (\partial_\rho L)^2 + {R^4 \eta_{\mu \nu} \over (L^2+\rho^2)^2} \partial^\mu L \partial^\nu L} 
\end{equation}
here $\rho$ is the radial direction corresponding to RG scale for the quark physics and $\mu,\nu$ run over the  3+1 spacetime directions. $\phi$ is the dilaton field of the background geometry that, by giving it $\rho, L$ dependence,
was used in \cite{Alvares:2012kr} to generate non-trivial dynamics.   The scalar $L$ in the D3/D7 system describes the position of the D7 brane in the AdS space
but it can also be considered simply as a scalar field dual to the quark mass and condensate. Indeed, near the 
chiral transition, $L$ and its derivatives are small and we can expand. To describe the vacuum we assume there is no space-time dependence of the quark fields and we have very simply (for constant $\phi$) 
\begin{equation}
 S \sim \int d \rho {1 \over 2} \rho^3 (\partial_\rho L)^2.
\end{equation}
If we write $\psi = L/\rho$ and integrate by parts this is nothing but the standard action for a scalar in AdS$_5$ with mass squared of $-3$. Such a field is known to describe a field theory operator of dimension 3 and its dimension 1 source. 
Solving the Euler-Lagrange equations leads to
%\begin{equation} \partial_\rho( \rho^3 L' ) = 0,  ~~~ L = m + {c \over \rho^2},  \hspace{1cm}
$ \psi = {m/\rho} + {c/\rho^3}$.
%\end{equation}
The coefficient $m$ and $c$ are interpreted, respectively, as the quark mass and the quark condensate. 

Top down models of chiral symmetry breaking and other phenomenological models have been developed using the D3/D7 system
essentially by putting in a potential for $L$ through the dilaton field, $\phi$, in the DBI action. In \cite{Alvares:2012kr} the relation between a choice of dilaton and the effective radial dependent mass of the scalar $L$ was explicitly determined. For
us here though this is just background.

We will work with the phenomenological holographic action
\begin{equation} \label{simple}
 S \sim \int d \rho {1 \over 2} \left[ \rho^3 (\partial_\rho L)^2 + \rho \Delta m^2(\rho,L) ~ L^2 \right]
\end{equation}
We have simply shifted the mass squared of the scalar $L$ by a $\rho$ and $L$ dependent contribution that we will choose
by hand. The standard AdS/CFT relation tells us that the mass of the AdS scalar is related to the dimension of the operator by $m^2 = \Delta (\Delta-4)$ where here $m^2 = -3 + \Delta m^2(\rho)$. The equation of motion for $L$ becomes
\begin{equation} \label{eom}
 \partial_\rho( \rho^3 L' ) - \Delta m^2 \rho L - \frac{1}{2} \rho L^2 \frac{\partial \Delta m^2}{\partial L} = 0 
\end{equation}

One can check a few simple examples: If $\Delta m^2=-1$ the solution is $L=k/\rho$ and it describes a dimension two operator and its dimension 2 source. If $\Delta m^2=+3$ the solution is $L = k \rho + k'/\rho^3$ and the operator is dimension 4 with a dimensionless source. An instability will set in if the mass squared of the scalar falls below the Breitenlohner-Freedman bound of $m^2=-4$ \cite{Breitenlohner:1982jf}, i.e. $\Delta m^2 <-1$.

\section{The Chiral Transition}

To model the chiral transition we must make a choice for the form of $\Delta m^2$ in (\ref{simple}). To begin with, let us choose the simple form
\begin{equation} \label{first} \Delta m^2 = - \Delta_{IR} e^{- {\alpha \over \Delta_{IR}} \sqrt{\rho^2+L^2}  } .
\end{equation}
This ansatz is zero in the UV, where the model will have canonical scaling dimensions, but falls in the IR corresponding to $\gamma_m$ growing.
%%KT: I split the following sentence into two; check that the original meaning is not changed!
The functional form (\ref{first}) reflects that in the D3/D7 system $L$ is a position of the brane and hence naturally enters into the radial direction of AdS along with $\rho$. Equally, though, it is a particular choice that lowers the mass in the IR at small $\rho$ if $L$ is also small, so it will tend to favour the generation of $L(0)$ which we need for chiral symmetry breaking. Note that the ansatz (\ref{first}) is linear in the IR at small $\rho$ with gradient $\alpha$. 

\begin{figure}[]
\centering
%\hspace{-2mm}
\includegraphics[width=6.5cm]{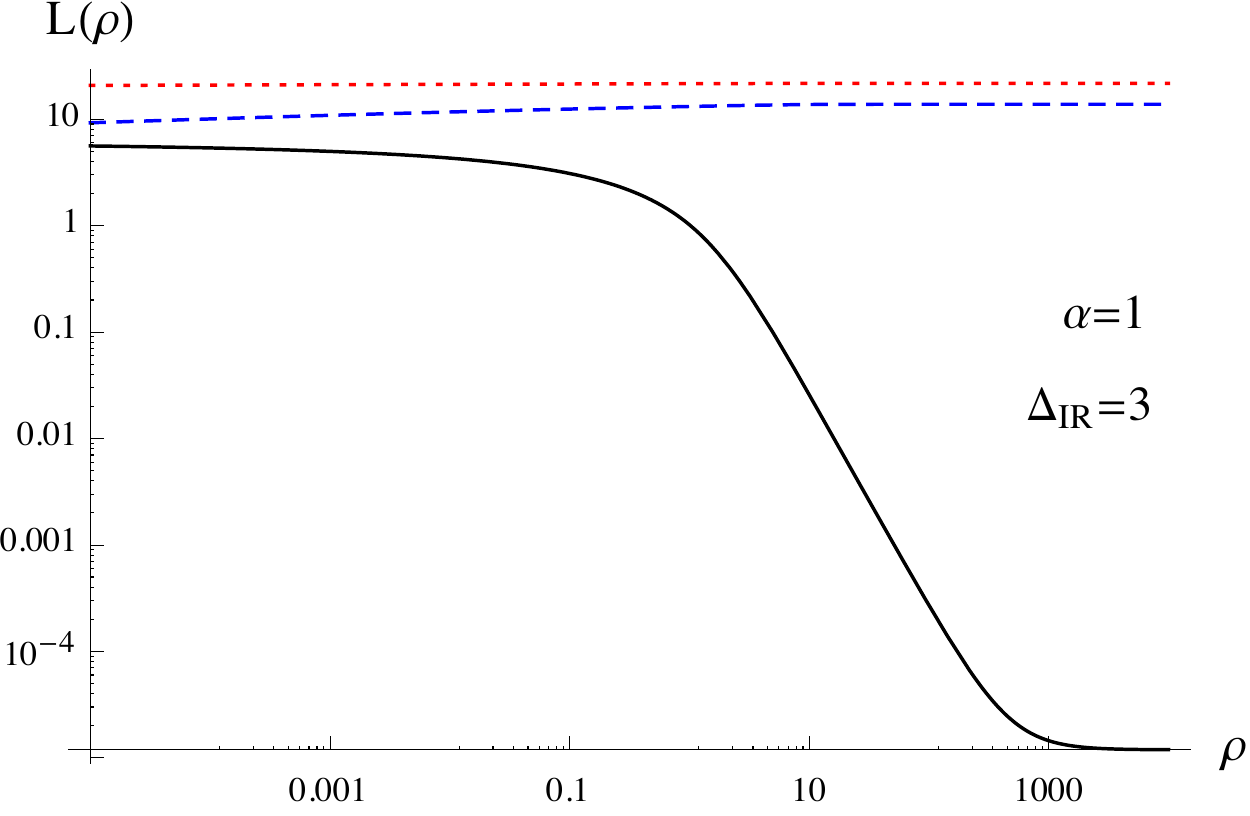}
\caption{Embeddings $L(\rho)$ for our model using the ansatz (\ref{first}) with  $\Delta_{IR}=3$ and $\alpha=1$. Red (dashed), blue (dotted) and black (solid) curves correspond to heavy, intermediate and zero quark mass, respectively.}
\label{L0embed}
\end{figure}

The values of the constants $\Delta_{IR}$, the IR value of $\Delta m^2$, and $\alpha$, which is the only scale in the problem (correponding to $\Lambda_{QCD}$), 
we imagine to depend on the running of the anomalous dimension of the quark bilinear operator due to the gauge dynamics. For example, if we take, arbitrarily, $\Delta_{IR}=3$ and $\alpha=1$ we can plot the solutions of the equation of motion (\ref{eom}). To pick a particular set of solutions we must require IR and UV boundary conditions. We choose
$L \rightarrow m$ in the UV and $\partial_\rho L=0$ in the IR - the latter is the natural choice in the D3/D7 system, and it seems a sensible choice in anycase. We interpret $L(0)$ as the IR value of the quark mass. We plot the embeddings, $L(\rho)$, in Fig. \ref{L0embed}. For large $m$ the solutions are flat.
At intermediate $m$ the IR dynamics somewhat reduces the quark self energy in the IR. There is no expectation of this in QCD but it is not a crucial aspect of any of the results below so we will not address it as a problem here. As $m$ approaches zero we see a positive IR mass generated, indicative of chiral symmetry breaking. Henceforth we will only consider massless UV solutions $m \rightarrow 0$.  

\begin{figure}[]
\centering
%\hspace{-2mm}
\includegraphics[width=6.5cm]{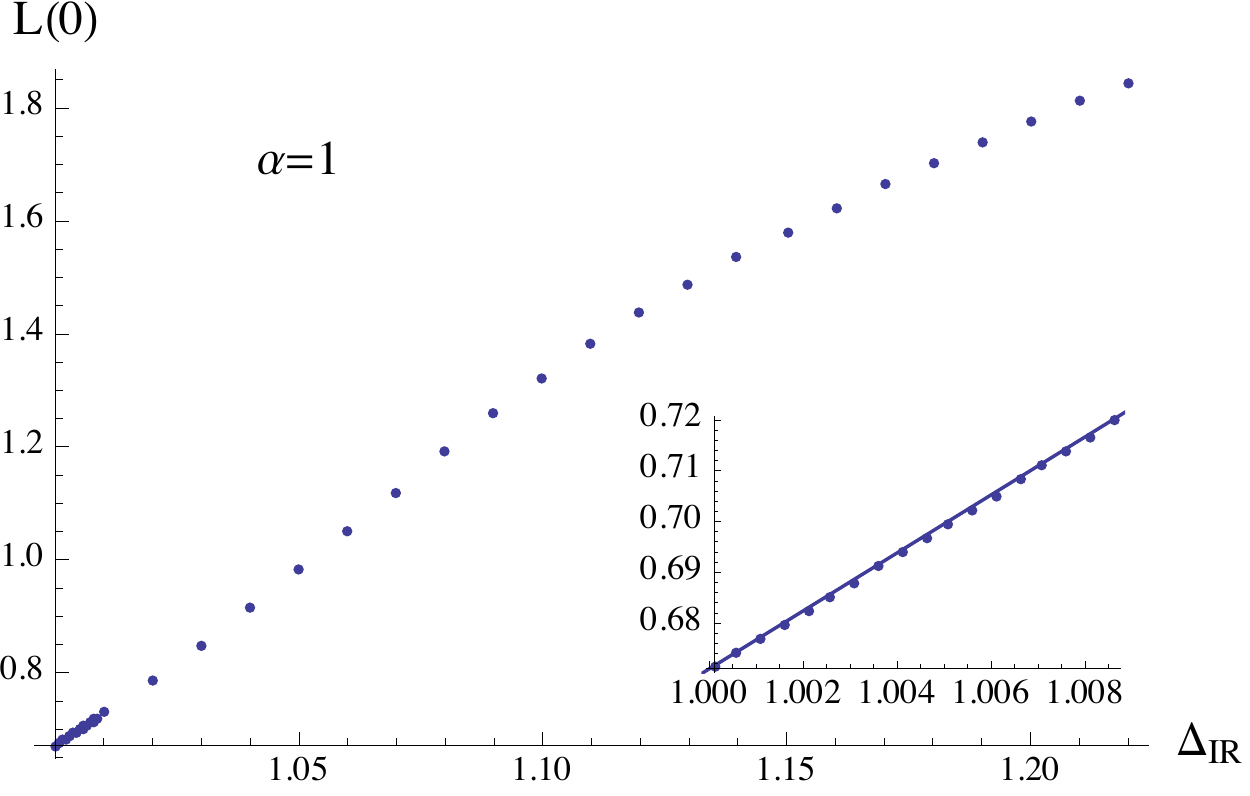}
\caption{ $L(0)$ against $\Delta_{IR}$ for the massless embeddings in our model using the ansatz (\ref{first})  with $\alpha=1$. The inset shows the numerical data at $\Delta_{\textrm{IR}}<1.01$ together with the linear fit $a+b (\Delta_{IR}-1)$ with $a=0.67$ and $b=5.74$.}
\label{L0values}
\end{figure}

The model can describe the chiral transition by simply dialling $\Delta_{IR}$ through $1$. At $\Delta_{IR}=1$ the BF bound is minimally violated at $\rho=L=0$ marking the onset of the transition. 
%As discussed in \cite{Alvares:2012kr}, the resulting transition is of the holographic BKT type with the condensate growing as $\exp(\Delta_{IR}-1)$ consistent with Miransky scaling. 
%% KT: revised text
We plot $L(0)$, the IR quark mass against $\Delta_{IR}$ in Fig. \ref{L0values}. As $\Delta_{IR}$ decreases towards $\Delta_{IR}=1$, the value of $L(0)$ falls linearly to a finite value at $\Delta_{IR}=1$ and then drops discontinuously to zero. Hence, the transition into conformal window is not of the BKT type, which would require $L(0)$ to fall exponentially towards zero, $L(0)\sim \exp(-1/(\Delta_{IR}-1))$, but a first order transition. 
%The fall below $\Delta_{IR}=1.1$ is fitted well by the expected exponential behaviour.

%%KT: Revised below
%We contend that the abrupt fall of $L(0)$ , and that it occurs when the BF bound is violated %(when naively using the AdS/CFT dictionary $\gamma_m=1$), are the two clear AdS predictions for %the transition into the conformal window. 
We contend that the prediction that the chiral transition occurs when the BF bound is violated (when naively using the AdS/CFT dictionary $\gamma_m=1$), is the clear AdS prediction for the transition into the conformal window. The model builder is then free to try to associate the point where the BF bound is violated to a specific value of
$N_f$ but in the absence of the true dual to these gauge theories it is necessarily speculative. Typically such models give $N_f^c/N_c \sim 4$ \cite{Jarvinen:2011qe,Alvares:2012kr}. The order of the transition depends on the gradient of the running in the IR - in this model we have a clear non-zero gradient and see a first order transition. If the conformal window exists then one instead expects a  vanishing 
gradient as one approaches the IR fixed point which will lead to a BKT transition. Our second model below (\ref{2scale}) will incorporate this additional key element. 

\begin{figure}[]
\centering
%\hspace{-2mm}
\includegraphics[width=6.5cm]{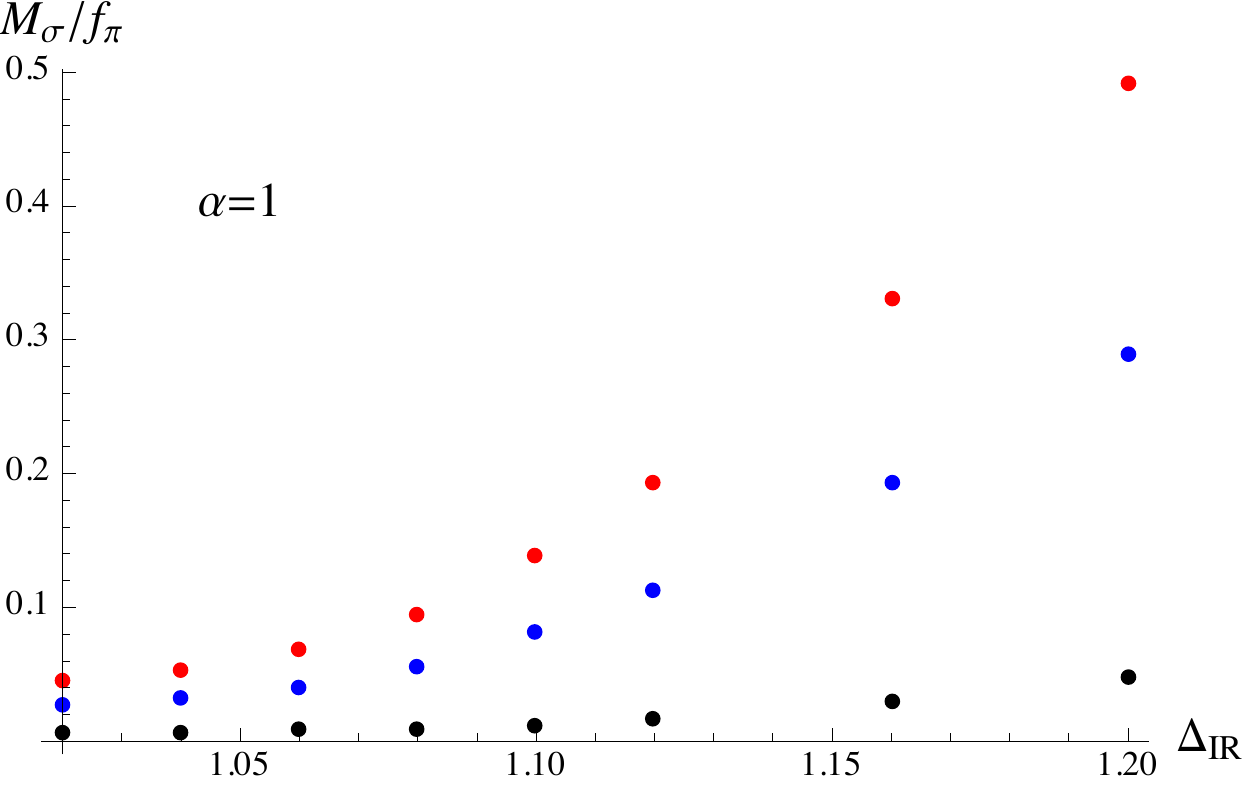}
\caption{The meson masses/$f_\pi$ against $\Delta_{IR}$ for the massless embeddings in our model using the ansatz (\ref{first}) and choosing $\alpha=1$.  }
\label{spectvsdelta}
\end{figure}

\section{Meson masses}

Let us now turn to computing the scalar $\bar{q}q$ meson masses of the model using the ansatz (\ref{first}). We will now be looking for space-time dependent excitations on top of the vacuum configuration, and must add the spatial derivatives of $L$.
We again bring over the linearized term from the DBI action of the D3/D7 system \cite{Karch:2002sh,Babington:2003vm,Alvares:2012kr} 
The Lagrangian has an extra term 
\begin{equation} \Delta {\cal L} = \frac{1}{2} \frac{\rho^3}{(L^2 + \rho^2)^2} R^4 (\partial_x L)^2 .\end{equation}
We write $L= L_0 + \delta (\rho) e^{ikx}$, $k^2=-M^2$, giving the equation of motion for $\delta$ as
\begin{equation} \label{deleom} \begin{array}{c}\partial_\rho( \rho^3 \delta' ) - \Delta m^2 \rho \delta -  2 \rho L_0 \delta \left. \frac{\partial \Delta m^2}{\partial L} \right|_{L_0} \\ \\
-  {1\over 2} \delta \rho L_0^2 \left. \frac{\partial^2 \Delta m^2}{\partial L^2} \right|_{L_0} 
+ M^2 R^4 \frac{\rho^3}{(L_0^2 + \rho^2)^2} \delta  = 0 \end{array} \end{equation}
We seek solutions with in the UV $\delta=1/\rho^2$ and with $\delta'(0)=0$ in the IR, giving a discrete meson spectrum. 
We can now for example compute the masses of the lightest two meson for the massless quark case in Fig 2. We will express our masses as ratios to the dynamically generated constituent quark mass $L(0)$. Note that formally we are computing 
$M^2 R^4$ with $R$ contributing a fixed numerical multiplier which will depend on the dynamics - our main result below is for the scaling of the masses rather than the absolute values, though. We expect, using naive dimensional analysis \cite{Georgi:1992dw}, that the pion decay constant will be given by
\begin{equation} \label{fpi} f_\pi^2 = {L(0)^2 \over 4 \pi^2} \end{equation}
For this case we find  $m_\sigma/f_\pi = 1.85$, $m_{\sigma^*}/f_\pi = 5.64$. These are somewhat low (by a factor of maybe 4) relative to QCD, but show that this example is in the right ball park to describe a natural gauge theory.

We can compute the dependence of the spectrum on $\Delta_{IR}$ to see how it varies as we approach the transition at $\Delta_{IR}=1$. Note that the $\alpha$ dependence is trivial since it is the only scale in the problem: Formally one can make the conformal transformation $L, \rho \rightarrow {L \over \alpha}, {\rho \over \alpha}, x \rightarrow \alpha x$ to set $\alpha=1$. Under this scaling $L(0) \rightarrow {L(0) \over \alpha}$, $M^2 \rightarrow {M^2 \over \alpha^2}$. The dimensionless ratio $M \over L(0)$ is therefore independent of $\alpha$. 

%%KT: I revised the discussion on the masses/goldstone spectrum
In Fig. \ref{spectvsdelta}  we show the spectrum for the lightest few meson states at $\alpha=1$ as we change 
$\Delta_{IR}$ through the transition into the conformal window. We see that the spectrum does become considerably lighter (relative to the natural value of one) as one approaches the transition, 
but there is no sense in which they are Goldstones whose masses move to zero there - they asymptote to fixed values at $\Delta_{IR}=1$. This is no surprise. First, in the IR  $\Delta m^2 \sim -\Delta_{IR} + \alpha \rho$ i.e. the gradient of the running is $\alpha=1$, and the theory is in no sense conformal. Second, the mass scale is given by the condensate, i.e. $L(0)$, which is finite all the way to the transition; hence the spectrum is expected to look roughly "QCD-like" towards the conformal window. A decrease in the meson masses by a factor of even 10-100 as we approach the BKT transition is potentially phenomenologically interesting. We will next consider a model that displays actual Goldstone behaviour. 

\begin{figure}[]
\centering
%\hspace{-2mm}
\includegraphics[width=6.5cm]{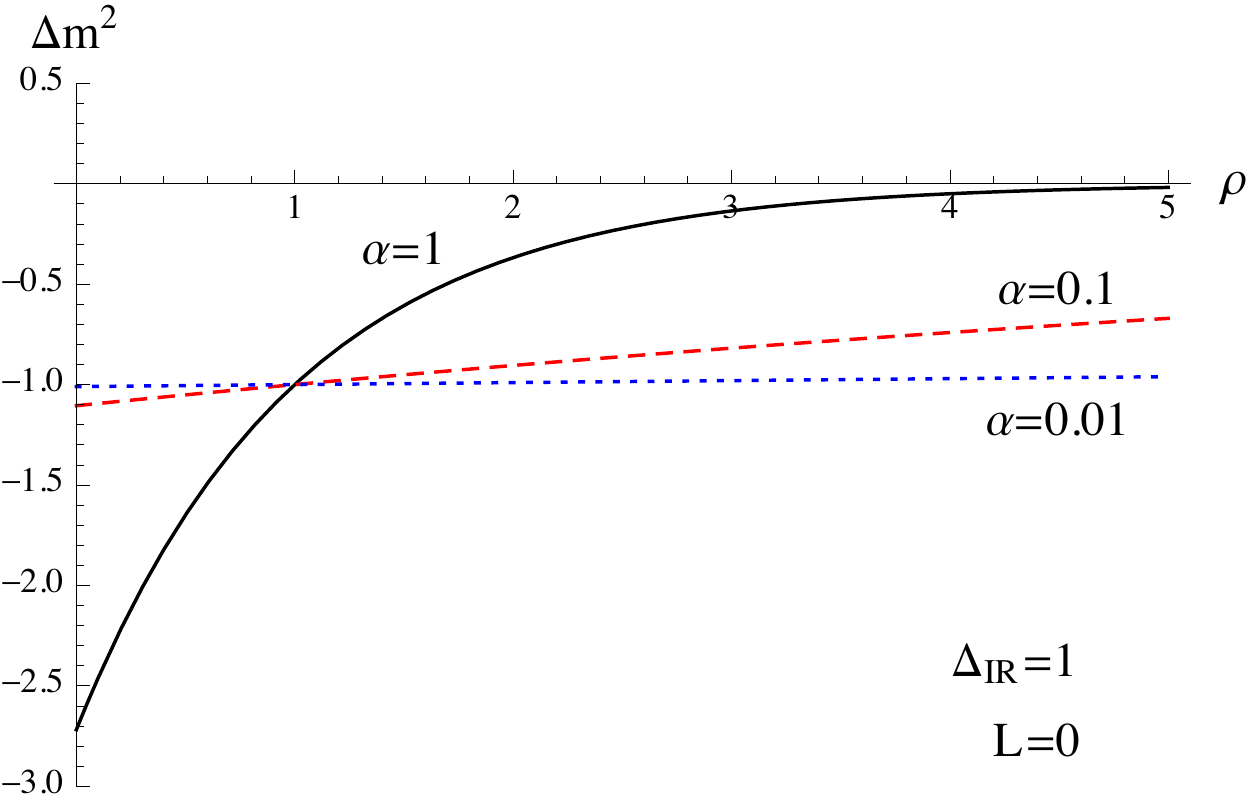}
\caption{Plot of the coupling ansatz in (\ref{2scale}) vs $\rho$ at $L=0$. The black (solid), red (dashed) and blue (dotted) curves correspond to $\alpha=1$, $0.1$ and $0.01$, respectively. }
\label{2scalepic}
\end{figure}

\begin{figure}[]
\centering
%\hspace{-2mm}
\includegraphics[width=6.5cm]{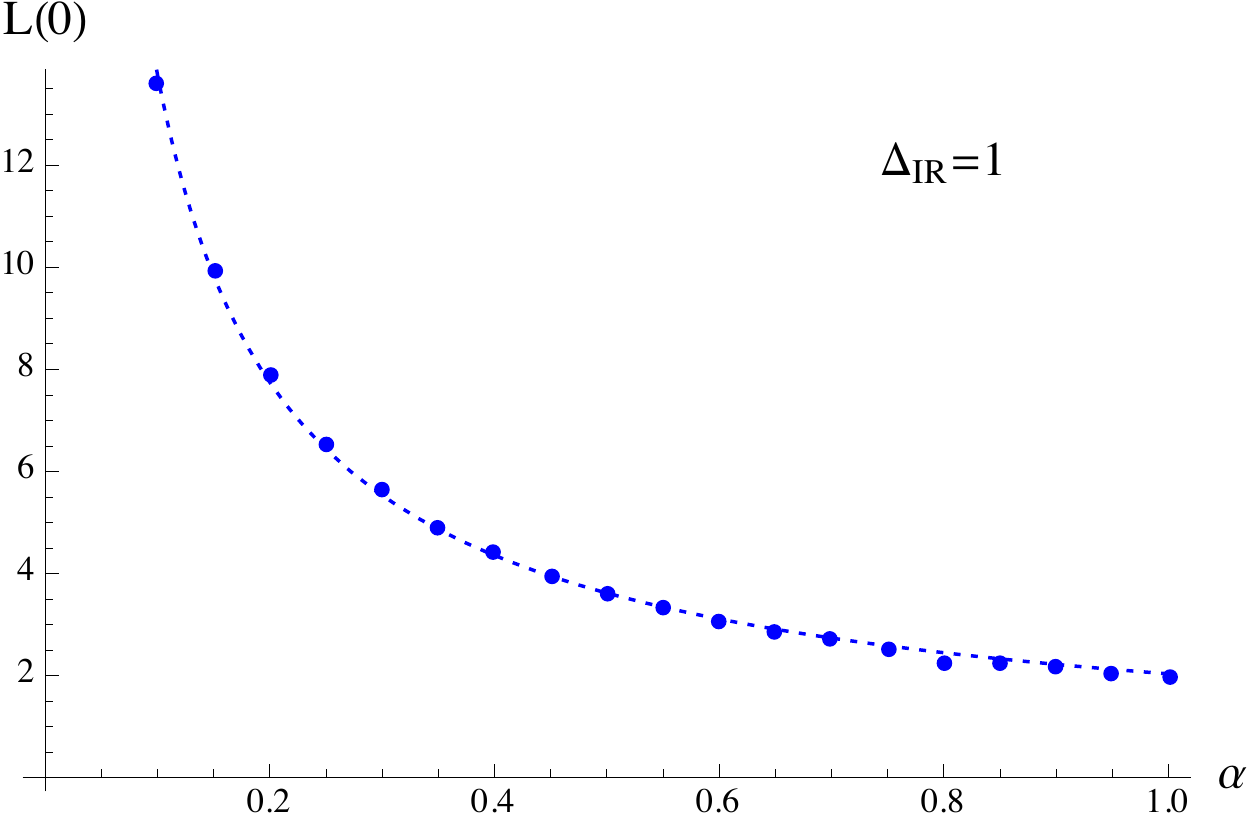}
\caption{Plot of L(0) for the massless embeddings vs $\alpha$ in (\ref{2scale}). $\Lambda=1, \Delta_{IR}=1$. The effective slope (shown by the dotted curve) is $\alpha^{0.83}$. }
\label{Lgrow}
\end{figure}

\section{Conformality and a Techni-Dilaton}

\begin{figure}[]
\centering
%\hspace{-2mm}
\includegraphics[width=6.5cm]{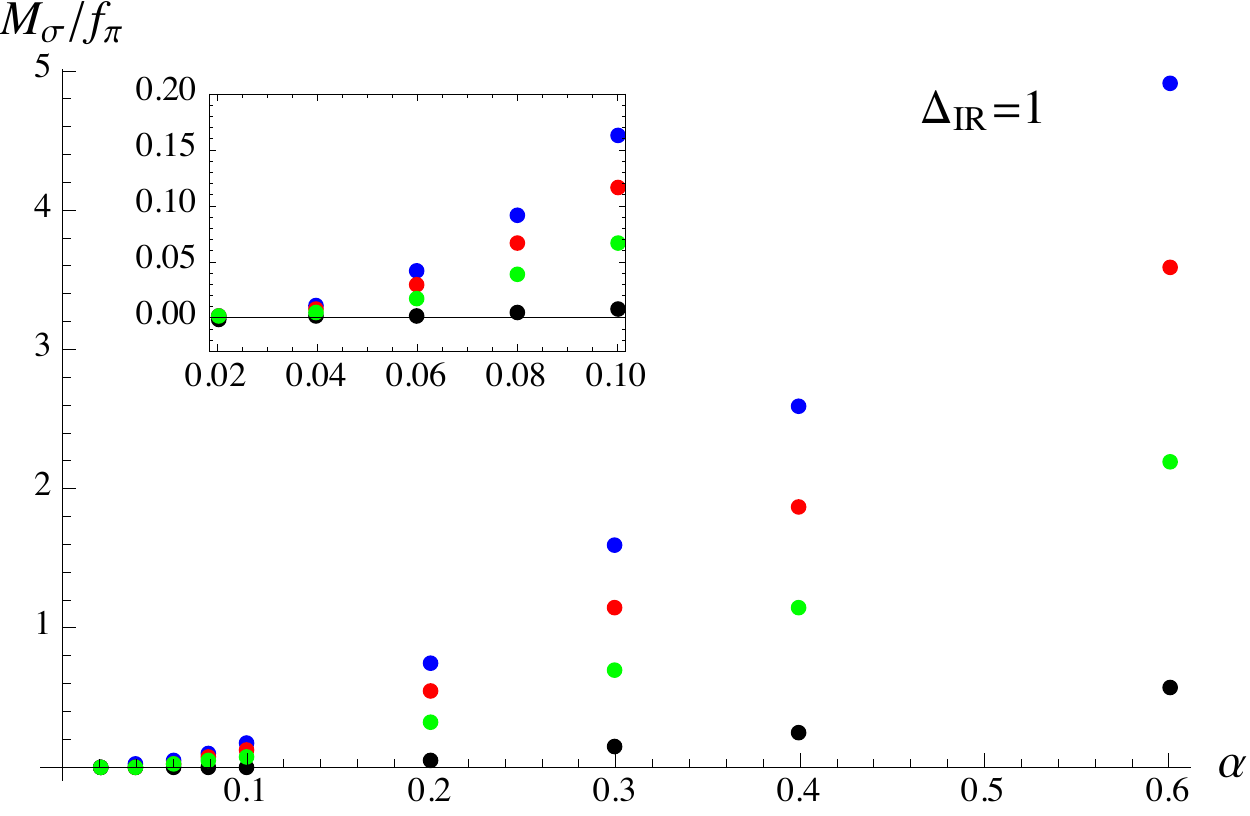}
\caption{The meson masses/$f_\pi$ vs $\alpha$  in (\ref{2scale}) for $\Delta_{IR}=1$.}
\label{mesmas}
\end{figure}

\begin{figure}[]
\centering
%\hspace{-2mm}
\includegraphics[width=6.5cm]{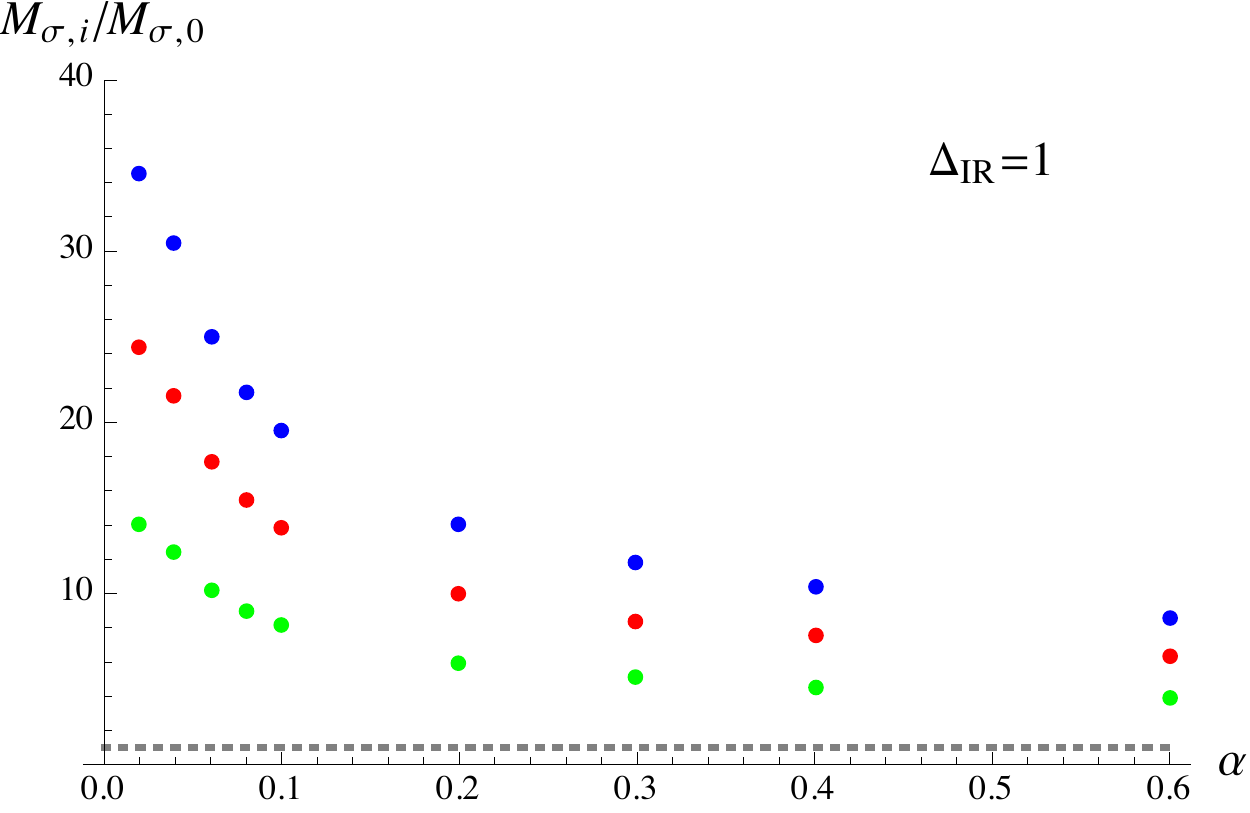}
\caption{The ratio of the excited meson masses to the $\sigma$ mass vs $\alpha$ in (\ref{2scale}) at  $\Delta_{IR}=1$.}
\label{ratio}
\end{figure}

We can now turn to the main question we wish to address in this paper. How does the mesonic spectrum change as we push our chiral symmetry breaking dynamics towards conformality? To address this we need a two scale ansatz for $\Delta m^2$ in Eq. (\ref{simple}); one scale to represent the intrinsic symmetry breaking scale and another the gradient of the running at that scale. We choose
\begin{equation} \label{2scale} \Delta m^2 = - \Delta_{IR} e^{- {\alpha \over \Delta_{IR}} \left( \sqrt{\rho^2 + L^2} - \Lambda\right)} \end{equation}
Again, in the far UV $\Delta m^2=0$ and the quark condensate has canonical dimension. Then, towards  the IR, $\gamma_m$ grows. Here $\Lambda$ is the scale at which $\Delta m^2= -\Delta_{IR}$. Since the symmetry breaking is triggered around $\Delta m^2=-1$ we will make the choice $\Delta_{IR}=1$ henceforth. While other choices can be made, the key questions are dependent on the gradient at the scale where 
$\Delta m^2=-1$, so fixing that scale is natural. We will also use a scale transformation to set the scale $\Lambda=1$ in our numerical studies and then the other dimensionful parameter $\alpha$ is given in units of $\Lambda$. The ansatz (\ref{2scale}) is such that the gradient of the running of $\Delta m^2$ is linear at the scale $\Lambda$ and given by $\alpha$ - see Fig (\ref{2scalepic}). All these theories break chiral symmetry since the BF bound is violated at scale $\Lambda$, but the theories become more conformal as $\alpha \rightarrow 0$. 
%%KT: Revised the text below.
Therefore $\alpha=0$ also corresponds to the transition into the conformal window, and we expect that the resulting transition is of the holographic BKT type with the condensate growing as $\exp(-1/(\Delta_{IR}-1))$ consistent with Miransky scaling \cite{Alvares:2012kr}. Thus $\alpha$ does three jobs: moves us towards the BKT transition, reduces the gradient of the running and increases the range over which the linear running holds. 

Firstly we can study the dependence of the embeddings on the gradient $\alpha$ by solving (\ref{eom}). Naively the scale of the BF bound violation is the same in all cases but the range of scale over which the BF bound is almost broken grows and this tends to enhance $L(0)$ as $\alpha \rightarrow 0$. We show this effect in Fig \ref{Lgrow}. Next, we can compute the $\sigma$ meson plus radially excited states' masses as a function of $\alpha$. We again solve (\ref{deleom}) numerically and express the meson masses in units of 
$f_\pi$ as defined in (\ref{fpi}).  The key result is that the full tower of states fall to zero mass as $\alpha \rightarrow 0$.   We show the results in the near BKT, near conformal, small $\alpha$ regime  in Fig. \ref{mesmas}. As $\alpha$ varies from 1 to 0.1 the drop is by roughly two orders of magnitude.   This is quite a fast drop off. The reason for the sharp dependence on $\alpha$ is, we believe, its multiple role as discussed above: as $\alpha$ decreases we simultaneously move towards the BKT transition, reduce the gradient of the running and increase the region of the linear running. Each of these effects seems to place the theory closer to conformality. Note that the drop in meson masses is not just the result of the increasing $L(0)$ as $\alpha$ decreases. The meson masses drop relative to the fixed scale $\lambda=1$ as well.

Finally in Fig \ref{ratio} we plot the ratio of the meson masses to the lightest $\sigma$ mass, and we see that the gap grows between the lightest state and the rest of the spectrum. This suggests it is reasonable to interpret the lightest $\sigma$ bound state as a pseudo-Goldstone boson associated with the spontaneous breaking of dilatation symmetry.  

\section{Phenomenological Implications}

We have presented a simple model of the scalar quark anti-quark state spectrum in a strongly
coupled gauge theory as the dynamics approaches a conformal nature. We have shown that the
mass of the lightest state scales to zero, relative to the dynamically generated quark mass,  as the gradient
of the running of the anomalous dimension falls to zero at the scale where $\gamma_m=1$ (or where the BF bound is violated in AdS). The lightest $\sigma$ particle also becomes light relative to the excited states with the same quantum numbers suggesting an interpretation of it as a pseudo-Goldstone associated to the spontaneous breaking of scale symmetry.  We have also seen that the masses of the higher radial excitations  scale to zero relative to the quark mass as $\alpha \rightarrow 0$ as they 
prepare to complete
a continuous spectrum at the conformal point.

Although our model is very basic, we should consider its implications in light of the recent discovery of a light higgs-like state at LHC \cite{atlas:2012gk,cms:2012gu}. Is it possible that this state is the $\sigma$ mode of some technicolour dynamics? Our results suggest that it could be if the dynamics is tuned towards
conformality. In QCD $m_\sigma / f_\pi \sim 6$ so in technicolour to obtain $m_h / v \sim 1/2$
is roughly a factor of 10 suppression. We would need to tune the gradient $\alpha$ in our model at roughly the 10$\%$ level to achieve this. To construct a full model, which is beyond our simple modelling here, one would also need to consider the size of the techni-dilaton decay constant \cite{Matsuzaki:2012xx} and the effects on its mass from the top quark yukawa coupling \cite{Foadi:2012bb}.

Our model suggests that other techni-quark bound states will also be anomalously light 
and one should expect to find additional scalar bound states, for example, starting at $\sim$10 times
the mass of the higgs. Such states are not
yet clearly experimentally ruled out, their precise masses will be dependent on the 
dynamics and we do not pretend to compute them exactly within our simple model. 

Let us consider walking near the lower boundary of the conformal window of SU($N_c$) gauge theory. A priori tuning to $\gamma \sim 1$ at the chiral symmetry breaking transition
and tuning the gradient of the coupling at the transition point are two separate tunings. The two are linked by
the beta function of the theory. Although the running of the theory at strong coupling is a matter of conjecture, there are some features which are scheme independent and robust, like the value of the anomalous dimensions or the slope of the beta function at the fixed point. 

As and example, consider the two-loop beta function for the 't Hooft coupling, $\lambda = g^2 N_c$, given at large $N_c$ and $N_f$ by  
\begin{equation}
\mu { d \lambda \over d \mu} =  - b_0 \lambda^2 + b_1 \lambda^3 \,,
\end{equation}
where
\begin{equation} 
b_0 = {2 \over 3} {(11 - 2 x) \over (4 \pi)^2}, \hspace{0.5cm}    
b_1 = -{2 \over 3} {(34 - 13 x) \over (4 \pi)^4} \,.
\end{equation}
If we assume chiral symmetry breaking sets in at $\gamma_m=1$,  and use the one loop result for the anomalous dimension then one finds the critcal coupling to be $\lambda_c= 8 \pi^2 / 3$. At $x=4$ this coincides with the fixed point value of the two loop running and gives an estimate of the location of the lower boundary of the conformal window.  

To estimate the degree of conformality as a function of $x=N_f/N_c$, one can look at how the gradient of the running coupling approaches the fixed point at $\lambda_c= 8 \pi^2 / 3$. For one loop beta function one finds
\begin{equation} \mu { d \lambda \over d \mu}=-b_1\lambda^2(\frac{b_0}{b_1}-\lambda)\sim\lambda_c^2(\lambda^\ast-\lambda_c)\sim (x-4) \end{equation}
near the lower boundary of the conformal window.
In other words the gradient of the running coupling is simply proportional to how close $x$ is to the critical value. For the observed higgs mass our model suggests one would need $x - x_c \sim 0.1$. 
\footnote{As a final remark, we note that the existence of Miransky scaling, i.e. a walking region directly below the lower boundary of the conformal window requires that the zero of the beta function at the lower boundary is at least of second order \cite{Sannino:2012wy,Tuominen:2012qu}. This gives a scheme independent characterization, and may affect the estimate we obtained above.}

The fine tunings we have discussed above are only mildly distasteful, but are inherent in a model with a light higgs and a gap
to the scale of any other new physics. At least the tuning of walking dynamics seems capable of generating a light higgs like state in combination with other benefical features such as the decoupling of flavour physics. 

\noindent {\bf Acknowledgements:} NE's work is supported by a STFC consolidated grant. We thank Matti Jarvinen for insightful comments on our first manuscript version.

\end{document}